\documentclass{article}

\usepackage{arxiv}

\usepackage[utf8]{inputenc} % allow utf-8 input
\usepackage[T1]{fontenc}    % use 8-bit T1 fonts
\usepackage{hyperref}       % hyperlinks
\usepackage{url}            % simple URL typesetting
\usepackage{booktabs}       % professional-quality tables
\usepackage{amsfonts}       % blackboard math symbols
\usepackage{nicefrac}       % compact symbols for 1/2, etc.
\usepackage{microtype}      % microtypography
\usepackage{lipsum}		% Can be removed after putting your text content
\usepackage{graphicx}
\usepackage{natbib}
\usepackage{doi}

\usepackage{amsmath}
\usepackage{amssymb}
\usepackage{xfrac}
\usepackage{dcolumn}% Align table columns on decimal point
\usepackage{array}

\title{Channel Formation Enhances Target Consumption by Chemotactic Active Brownian Particles}

\author{ \href{https://orcid.org/0000-0001-7930-9622}{\includegraphics[scale=0.06]{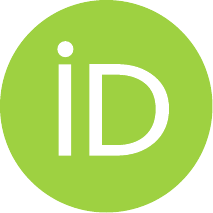}\hspace{1mm}Vladimir Yu.~Rudyak} \\
	School of Physics and Astronomy\\
    Tel-Aviv University\\
    Tel Aviv, Israel 69978
	\texttt{vurdizm@gmail.com} \\
	\And
	\href{https://orcid.org/0000-0000-0000-0000}{\includegraphics[scale=0.06]{orcid.pdf}\hspace{1mm}Shahar Shinehorn} \\
	School of Chemistry\\
    Tel-Aviv University\\
    Tel Aviv, Israel 69978
	\texttt{shinehorn@mail.tau.ac.il} \\
    \And
	\href{https://orcid.org/0000-0003-1927-4506}{\includegraphics[scale=0.06]{orcid.pdf}\hspace{1mm}Yael Roichman} \\
    School of Physics and Astronomy \&\\
	School of Chemistry\\
    Tel-Aviv University\\
    Tel Aviv, Israel 69978
	\texttt{roichman@tauex.tau.ac.il} \\
}

\renewcommand{\shorttitle}{Channel Formation Enhances Target Consumption by Chemotactic ABPs}

\hypersetup{
pdftitle={A template for the arxiv style},
pdfsubject={q-bio.NC, q-bio.QM},
pdfauthor={David S.~Hippocampus, Elias D.~Striatum},
pdfkeywords={First keyword, Second keyword, More},
}

\begin{document}
\maketitle

\begin{abstract}
In many situations, simply finding a target during a search is not enough. It is equally important to be able to return to that target repeatedly or to enable a larger community to locate and utilize it. While first passage time is commonly used to measure search success, relatively little is known about increasing the average rate of target encounters over time. Here, using an active Brownian particle model with chemotaxis, we demonstrate that when a searcher has no memory and there is no communication among multiple searchers, encoding information about the target’s location in the environment outperforms purely memoryless strategies by boosting the overall hit rate.  We further show that this approach reduces the impact of target size on a successful search and increases the total utilization time of the target.
\end{abstract}

% keywords can be removed

rom the abstract domain of computer science, where algorithms systematically navigate solution spaces, to the complex behaviors of living organisms seeking food and safety, the pursuit of a target is a universal phenomenon. Over the years, this search problem has been studied from a variety of perspectives. Some strategies focus on enhancing the search algorithm itself by incorporating information \cite{hart1968formal}, memory \cite{mnih2015human}, or even principles of natural selection and machine learning \cite{silver2016mastering,sutton2018reinforcement}. Other approaches draw inspiration from collective animal behavior, such as the flocking of birds or schooling of fish \cite{kennedy1995particle}. Still others optimize the searcher's mode of motion -- for example, the intermittent movement patterns of predatory birds \cite{Benichou2011}, or the stochastic returns to the hive observed in foraging insects \cite{evans_stochastic_2020}.
Generally, the success of a search strategy is measured by the mean time it takes the searcher to find the target for the first time, i.e., the mean first passage time (MFPT), focusing on the initial discovery \cite{TargetSearchProblems2024, Redner2001}. However, while some search objectives require only initial discovery, others demand repeated target visits. Ant colonies exemplify the latter, as scouts locate large food sources that then require coordinated retrieval by multiple foragers \cite{holldobler1990ants,jeanson2007flexibility,dussutour2009ants}. 

{Chemotaxis serves as a crucial coordination mechanism and means of indirect communication among decentralized biological agents when direct methods are unavailable \cite{DETRAIN2006,Perna2012,czaczkes2015trail}. This is evident in social insects like ants, which utilize pheromone trails to create self-organized structures that guide colony members to resources \cite{DETRAIN2006,Perna2012,czaczkes2015trail}. 
These structures emerge from simple individual behaviors, resulting in efficient foraging without centralized control \cite{Amorim2018, Kumar2019}. A key outcome in many chemotactic systems is the formation of narrow, preferential paths known as channels, facilitated by spatial persistence of agent motion, temporal persistence of environmental memory, and interaction rules that promote alignment with existing trails \cite{Caillerie2018, Feinerman2018, Amorim2021, Mok2023}. These channels introduce spatial anisotropies that depart from isotropic search dynamics \cite{Benichou2011,TargetSearchProblems2024} and play a significant role in directing agents toward search targets, ultimately enhancing transport and foreage efficiency. Similar phenomena occur in other biological systems, including bacterial colonies \cite{BenJacob2000}, slime molds \cite{Reid2012}, and even in abiotic active matter systems \cite{Altshuler2024,Romanczuk2012,Popescu2018,Hokmabad2022}, suggesting fundamental physical principles underlying these behaviors.}

Despite significant advances in understanding chemotaxis and environmental memory, it remains unclear in what manner channel formation optimizes target search. Specifically, MFPT fails to capture the benefits of gradual target consumption \cite{TargetSearchProblems2024, Redner2001}. Second, what are the minimal requirements for symmetry breaking that lead to directed channels toward targets? The relative contributions of persistence, alignment, and reinforcement mechanisms remain unclear \cite{Mok2023, Morin2015, Buerle2018}. This paper addresses these gaps by demonstrating that self-organized channels formed by chemotactic active Brownian particles optimize the hit rate for targets requiring multiple encounters and significantly reduce sensitivity to target size. Our findings reveal that environmental memory transforms the search paradigm from one optimized for initial discovery to one designed for efficient repeated comsumption.

%\textcolor{blue}{The GAP: what is the underlying requirement for symmetry breaking for channels towards targets, what is it optimizing? Answer - optimal hit rate and reduced sensitivity to target size }

\section{Chemotaxis leads to channeling of active Brownian particles}
%\textcolor{blue}{A simple model  - how channeling arises - have to say explicitly that we use a modified Kegel-Segal model that we modified to apply to ABPs and cite them and if there is another paper that does what we did (I think there is).}

\begin{figure*}[ht!]
\centering
\includegraphics[width=1.0\linewidth]{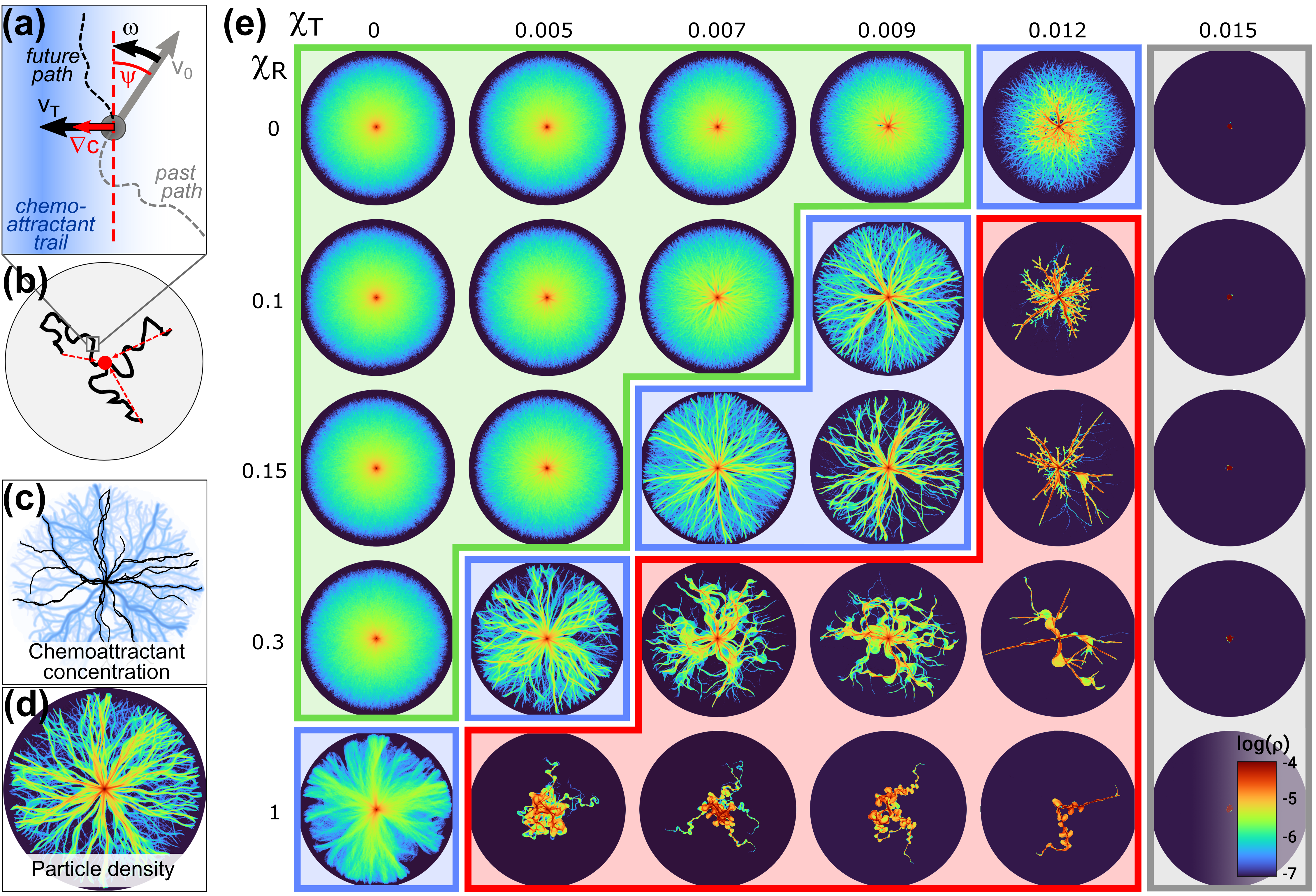}
\caption{\small{(a) Schematic representation of an active Brownian particle (ABP) with chemotaxis moving in the vicinity of a chemoattractant trail. $\mathbf{v}_0$ is the ABP default velocity, $\nabla c$ is the gradient of chemoattractant concentration, $\mathbf{v}_\text{T}$ is the velocity gained by attraction towards higher chemoattractant concentration, and $\omega$ is the ABP direction rotation towards the chemoattractant traverse direction (dashed line).
(b) Schematic representation of an arena and resetting process. (c) Instantaneous chemoattractant concentration field (orange) after $10^5$ simulation time in small arena ($R=l_p$), and next 20 particle trajectories (black lines). (d) Particle density $\rho$ after $10^5$ simulation time in small arena ($R=2l_p$). (e) Channeling regimes occurring at various translational $\chi_T$ and rotational $\chi_T$ chemosensitivities: weak channeling (green), strong channeling (blue), extreme channeling (red), and collapse (gray).
}}
\label{fig:channeling}
\end{figure*}
Our modeling approach to study search with environmental memory employs three ingredients: a searcher is a persistent self-propelling particle; it creates a pheromone (chemoattractant) trail as it walks and simultaneously interacts with it; 
and its search is also truncated via sharp resetting to avoid boundary effects.

Specifically, we consider an overdamped Active Brownian Particle (ABP) \cite{ebeling1999active,romanczuk2012active} with a fixed self-propulsion speed, $v_0 \equiv 1$, and purely rotational diffusion (i.e. no translational diffusion). We use Keller-Segel continuum reaction-diffusion description \cite{Keller1971a, Keller1971b} modified for particle-based simulations to model its interaction with the chemoattractant field, $c(\mathbf{r},t)$. As the particle moves, it deposits a chemoattractant with intensity $\beta$. It is attracted toward regions of high chemoattractant concentration, with its translational velocity aligned with the chemoattractant gradient ($\mathbf{v}_T \sim \nabla c$; see Fig.,\ref{fig:channeling}a). Simultaneously, its orientation vector $\mathbf{n}$ rotates perpendicularly to the chemoattractant gradient, effectively aligning with an inferred trail (illustrated as a red dashed line in Fig.~\ref{fig:channeling}a). The chemoattractant concentration decays exponentially over time, enabling the system to reach a non-trivial steady state. The search in our simulations is conducted sequentially, one searcher at a time, each interacting with the trails left by all previous searchers. This process is described by the following coupled equations,

\begin{align*}
&d\mathbf{r}(t) = v_0\mathbf{\hat u}(\mathbf{r},t)\,dt\\
&\mathbf{u}(\mathbf{r},t)=\chi_T\frac{\nabla c(\mathbf{r},t)}{(c(\mathbf{r},t)+c_0)}+\mathbf{n}(t)\\
&d\varphi(t) = \left(\chi_{R} \frac{|\nabla c(\mathbf{r},t)|\cdot \sin(2\psi(t))} {\left(c(\mathbf{r,t})+c_0\right)}+ \sqrt{2 D_{R}}\eta(t)\right)dt \\
& dc(\mathbf{x},t) = \left(-\frac{1}{\tau_{c}}c(\mathbf{x},t)+ \beta N_{0,r_c}(|\mathbf{x}-\mathbf{r}(t)|)\right)dt
\end{align*}

where $\mathbf{n}(t)=\{\cos\varphi(t);\sin\varphi(t)\}$ is the particle direction, $\psi(t)=\angle(\mathbf{n}(t),\nabla c(\mathbf{r},t))$ is the angle between the particle direction and chemoattractant gradient, $c_0$ is the concentration sensitivity noise level, $\chi_{R,T}$ are the rotational and translational chemosensitivities, $\eta(t)$ is a white noise, and $N_{0,r_c}(r)$ is a Gaussian profile with zero mean and standard deviation of $\sigma=r_c$.  For a detailed description of the model, see the Materials and Methods section. Similar particle-based simulations with chemotactic responses have successfully reproduced trail formation in ant colony-like systems and revealed universal features of collective movement across various systems \cite{Ryan2015, Amorim2021, Mok2023, Dodokov2024}.

Our simulations begin by placing a single particle in the center of a round pristine arena, $c(r,t=0)=0$, of radius $R$. The initial direction of the particle is chosen randomly, and its search duration is $\tau_R\equiv R/v_0$. After the allotted time, we reset the particle's position to the center of the arena and assign it a new random direction (Fig.~\ref{fig:channeling}b). Resets are considered instant, and the chemoattractant field remains unchanged during each reset. The particle is then released to continue its search. This procedure is repeated continuously, leading to the accumulation of the chemoattractant field and the emergence of characteristic features, such as channels with high chemoattractant concentration (Fig.~\ref{fig:channeling}c). The searching particles predominantly follow these channels (Fig.~\ref{fig:channeling}d).

The channels emerge through a stochastic process, meaning each independent simulation run yields a unique channel configuration. While the persistence of these channels arises from the collective interactions, it is critically dependent on the temporal stability of the environmental memory, governed by the chemoattractant decay time $\tau_{c}$. We define a non-equilibrium steady state when the total chemoattractant concentration stabilizes.  Stable channels form only when $\tau_{c}$ is significantly longer than the single trajectory time $\tau_{R}$. However, these channels are not permanent and their structure changes randomly at simulation times much greater than $\tau_{c}$ ($t_{sim} \gg \tau_{c}$). In all simulations, we set $\tau_{c}$ to correspond to the duration of 1000 trajectories.

To explore the range of behaviors in this system, we simulated a single particle with parameters $v_0 = 1$, $D_R = 1$, and an intrinsic persistence length $l_p^{(0)} = 1$, confined within a small circular arena of radius $R = 2l_p^{(0)}$. We systematically varied the translational and rotational chemosensitivities, $\chi_T$ and $\chi_R$, respectively. Each simulation consisted of $5 \times 10^4$ trajectories with a total simulation time of $t_\text{sim} = 10^5$, and statistical measurements were taken in the non-equilibrium steady state for $t > 0.5 \times 10^5$.

Figure~\ref{fig:channeling}e presents the resulting particle density profiles for selected parameter sets. We identify four distinct behavioral regimes: weak channeling (green), strong channeling (blue), extreme channeling (red), and collapse (gray). While the visual differences between these regimes are apparent, the classification was quantitatively determined via analysis of the trajectory persistence length ({see Supplementary Information}). In weak channeling regime, the searcher almost doesn't demonstrate preferrential paths. Increase of chemosinsitivity results in strong channeling, where channels are very persistent in time, but searcher still can avoid it and explores the whole arena. Importantly, both translational and rotational chemosensitivities can lead to the formation of channels. Further increase of chemosensitivity leads to the extreme channeling, when the searcher gets permanently trapped into emerging channels, and only a small part of the arena surface is explored. Here, translational sensitivity has a stronger impact, typically generating dense, branching networks of short-persistence channels. In contrast, rotational sensitivity favors the formation of straighter, non-branching channels that are broader and less focused. Finally, in the strong-attraction limit, this effect leads to particle localization near the origin, called collapse. We will henceforth focus on the strong channeling regime, as it resemble the structures observed in natural systems, such as the trail networks formed during ant colony searches.

\section{Channeling increases first passage times}

\begin{figure}[h!]
\centering
\includegraphics[width=.65\linewidth]{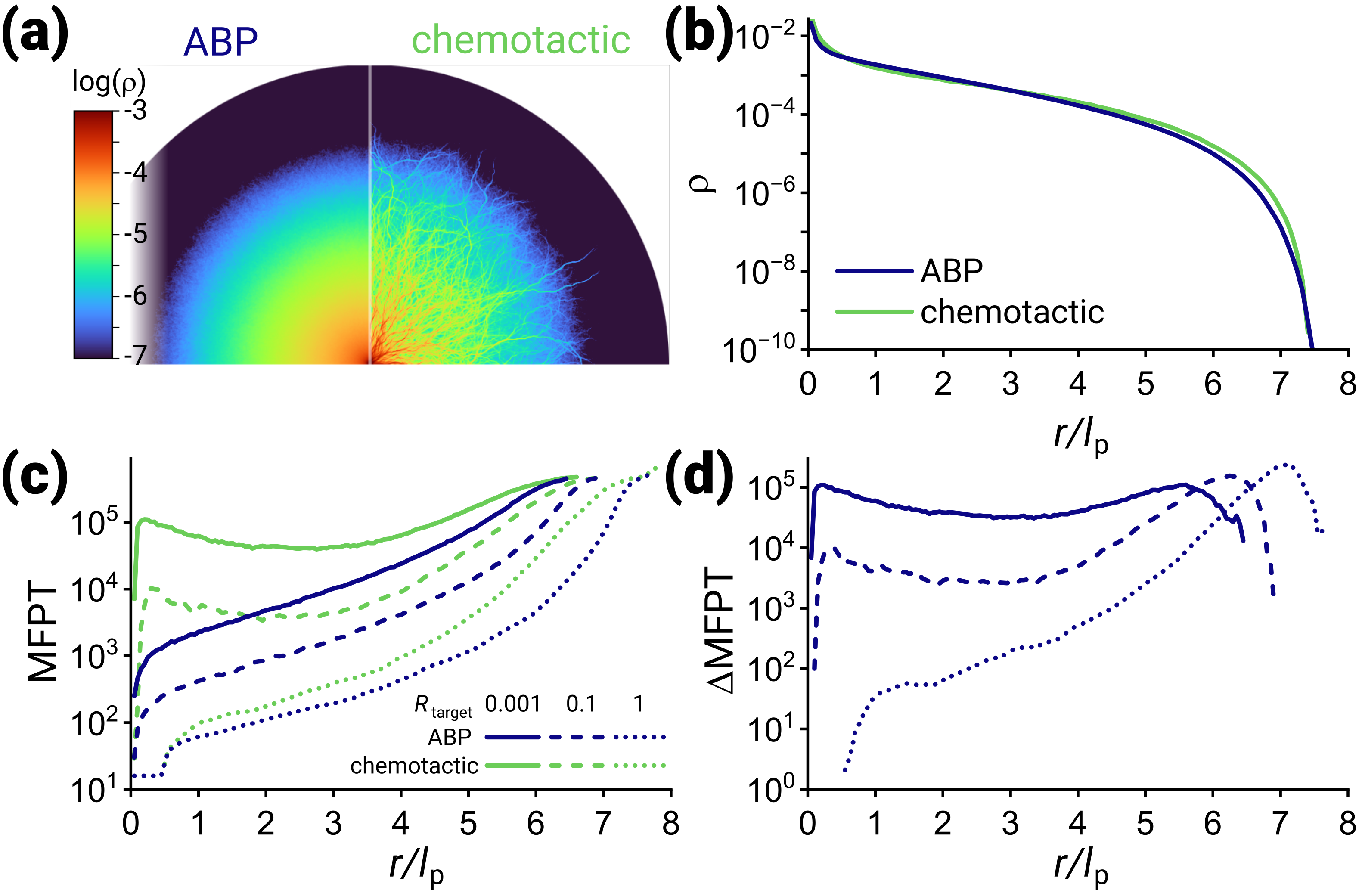}
\caption{\small{Comparison of particle density $\rho$ maps (a) and radial profiles (b) for pure active Brownian particles (ABPs, blue) and chamotactic particles (green) in a large arena $R=8l_p$. Mean first passage time dependencies on distance from the arena center to the target relative to persistence length, $r/l_p$, for various target radii $R_\text{target}$ (c). Mean first passage time difference between chemotactic and active Brownian particles, $\Delta\text{MFPT}$.}}
\label{fig:fpt}
\end{figure}

To assess the impact of environmental memory on search efficiency, we first examined the mean first-passage time for a searcher to reach a target, comparing systems with and without chemotaxis. Figure~\ref{fig:fpt}a shows representative particle concentration maps as a function of distance from the arena center for two cases: a pure active Brownian particle (ABP) without chemotaxis (left), and a chemotactic particle exhibiting strong channeling ($\chi_R = 0.15$, $\chi_T = 0.009$, right).
Simulations were conducted in a large arena ($R = 8 l_p$, $\tau_{c} = 1.6 \times 10^4$). Each result was averaged over 10 independent realizations, with a total simulation time of $t_\text{sim} = 10^6$ per realization (62,500 trajectories).
Since the persistence length of the searcher is strongly influenced by the chemoattractant field, we fixed it to $\ell_p = 2$ in both systems to ensure a fair comparison. This corresponds to rotational diffusion constants of $D_R = 1$ for the chemotactic system and $D_R = 0.5$ for the non-chemotactic one. As a result, the steady-state density profiles of both systems appear similar (Fig.~\ref{fig:fpt}b).

The mean first-passage time (MFPT) naturally depends on the target’s distance from the origin, $r$, and the target radius, $R_\text{target}$. For pure active Brownian particles (ABPs), the MFPT exhibits a characteristic exponential dependence on \( r \) for small targets~\cite{baouche2024optimal}. In density depletion regions, where \( r/\ell_p \gtrsim 5 \), this is followed by an even steeper increase (Fig.~\ref{fig:fpt}c, black lines). However, for chemotactic particles, the MFPT shows a non-monotonic dependence on distance (Fig.~\ref{fig:fpt}c, red lines), being surprisingly high for small targets near the origin. This counterintuitive result arises from the searcher's repeated, inefficient revisits within channels, which limit the explored area. Because channels most strongly attract particles near the origin, they impede the search most significantly in that region. At larger distances, the exploration redundancy drops, but is still strong enough to increase MFPT compared to pure ABP searcher (Fig.~\ref{fig:fpt}d). Therefore, in strong channeling conditions, chemotaxis is detrimental for searches targeted at first discovery. Yet, the prevalence of chemotactic channeling in natural systems indicates that it should provide some foraging benefit.

\section{Feedback speeds up consumption of gradually depletable targets}

In natural systems, channeling behavior typically emerges when colonies encounter targets that require multiple individuals or repeated visits for complete consumption, such as large food sources. 
Notably, insects are not the only good example. Some large animals also exhibit channeling (often called `game trails') to revisit critical persistent targets like water and mineral sources, while
behaving more individually against small or non-persistent ones. These include African savannah elephants \cite{Polansky2015, Stears2025}, followed by other species \cite{Nomoto2025}, and even some predators \cite{Blake2011}.
These ``heavy'' targets remain active after the initial contact and require a certain number of encounters required for complete consumption (called target weight, $W$). However, the MFPT only captures the time required for a searcher (or colony) to become \textit{aware} of the target’s location. For ``heavy'' targets, full consumption requires additional visits, and the appropriate measure becomes the \textit{mean first passage count} (MFPC), representing the mean number of arrivals at a target as a function of time.

In the absence of environmental memory, the average time to consume the target is simply given by $t_\text{cons} = W \times \text{MFPT}$. As shown above, channeling does not reduce the MFPT, and therefore cannot decrease $t_\text{cons}$. In contrast, environmental memory allows the system to encode the target's location by altering chemoattractant deposition upon reaching the target, effectively reinforcing the memory of its position.

\begin{figure}[htbp]
\centering
\includegraphics[width=.65\linewidth]{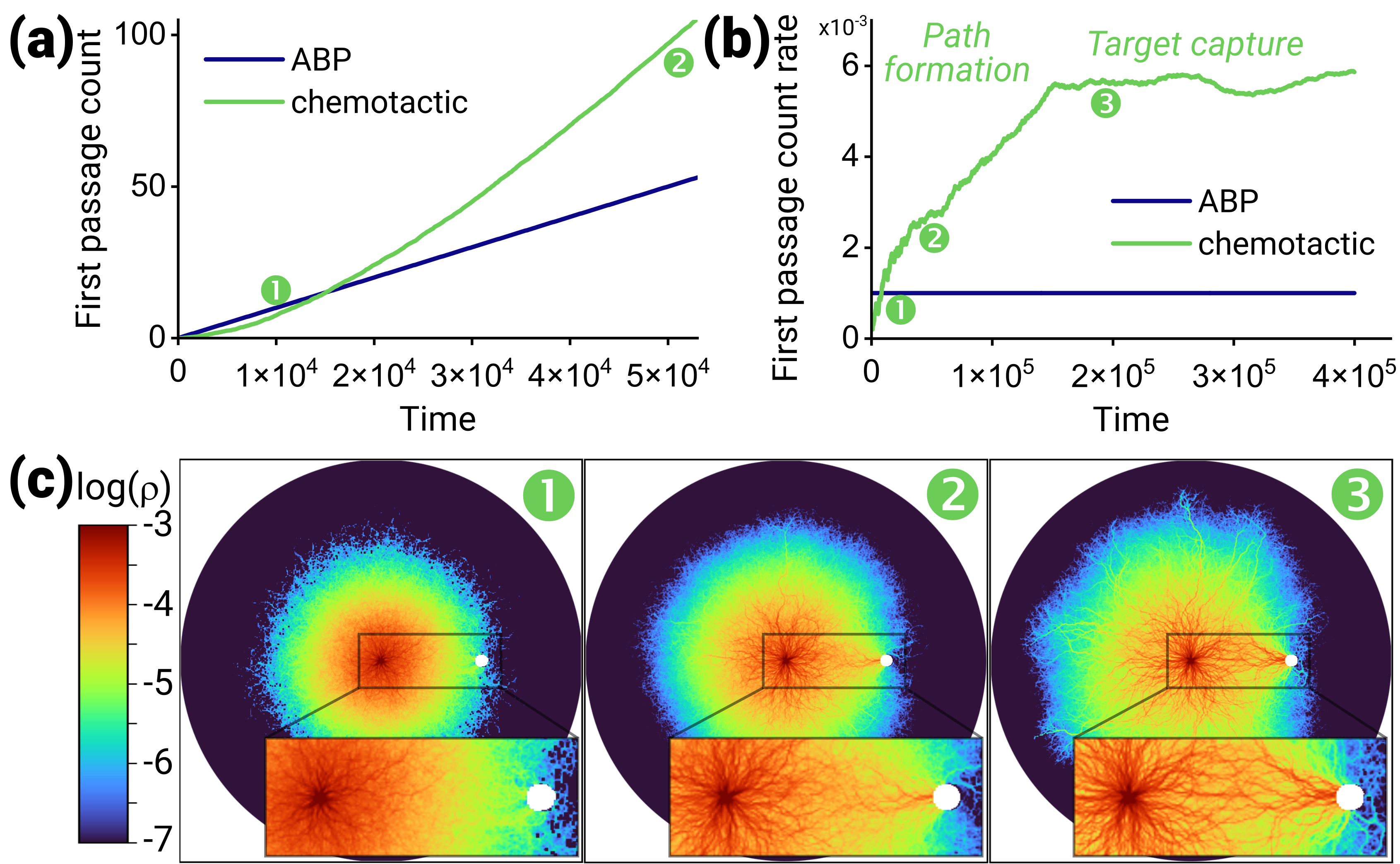}
\caption{\small{Target hits dynamics for pure active Brownian particles (ABPs) and chemotactic ones in large arena $R=8l_p$, distance to target $d=4l_p$, target radius $R_\text{target}=\nicefrac{1}{4}l_p$. Mean first passage count as a function of time (a). Mean first passage count rate (b). Cumulative particle densities $\rho$ (c) in the beginning (mark 1), during the channel formation (mark 2), and for established channels (mark 3), $t_\text{sim}=10^4$, $5\times10^4$, and $2\times10^5$, accordingly.}}
\label{fig:pathformation}
\end{figure}

To study the effect of memory reinforcement on target consumption, we set a small circular target of radius $R_\text{target} = \nicefrac{1}{4} l_p$ at a half-radius distance $d = 4 l_p$ from the center of the large circular arena ($R = 8 l_p$, $\tau_{c} = 1.6 \times 10^4$). We used strong target reinforcement ($\beta_\text{mult}=100$, see Target Reinforcement in the Materials and Methods section for details). We then compared the behavior of ABPs with and without chemotaxis. Figure ~\ref{fig:pathformation}a  shows the MFPC over time for both scenarios. Initially, standard ABPs exhibit more target hits, which aligns with their shorter MFPT due to lack of memory; their mean first passage count rate (MFPC rate) remains constant (Fig.~\ref{fig:pathformation}b). However, at later times, the chemotactic search becomes more effective, surpassing the performance of regular ABPs. This shift is evident in the chemoattractant field's evolution (Fig.~\ref{fig:pathformation}c), where dominant channels progressively form, focusing the broader network of channels toward the target. The development of this target-directed structure leads to an increasing MFPC rate, which eventually saturates to a constant value as the chemoattractant field reaches a steady state (Fig.~\ref{fig:pathformation}a, inset). Consequently, a system employing chemotaxis can consume heavy targets considerably faster than the one without it, as demonstrated in this case for $W>15$.

The analysis of the MFPC rate reveals three stages of target consumption: (i) target search, (ii) path formation (where the MFPC rate grows until the system reaches a steady state), and (iii) target capture at the established NESS rate. Fig.\,\ref{fig:pathformation}a shows that the significance of each stage is determined by the target weight. The dominating factor for targets with $W\leq10$ is the MFPT. For intermediate weight targets ($10\leq W \leq 100$), it is the path formation time, and for the heaviest targets ($W\geq 100$) it is the maximum target hit rate achieved in the steady state. We note that the distance to the target does not significantly influence this behavior until the target can be reached in principle ({see Supplementary Information}).

\section{Channeling reduces sensitivity to target size}

\begin{figure}[h]
\centering
\includegraphics[width=.50\linewidth]{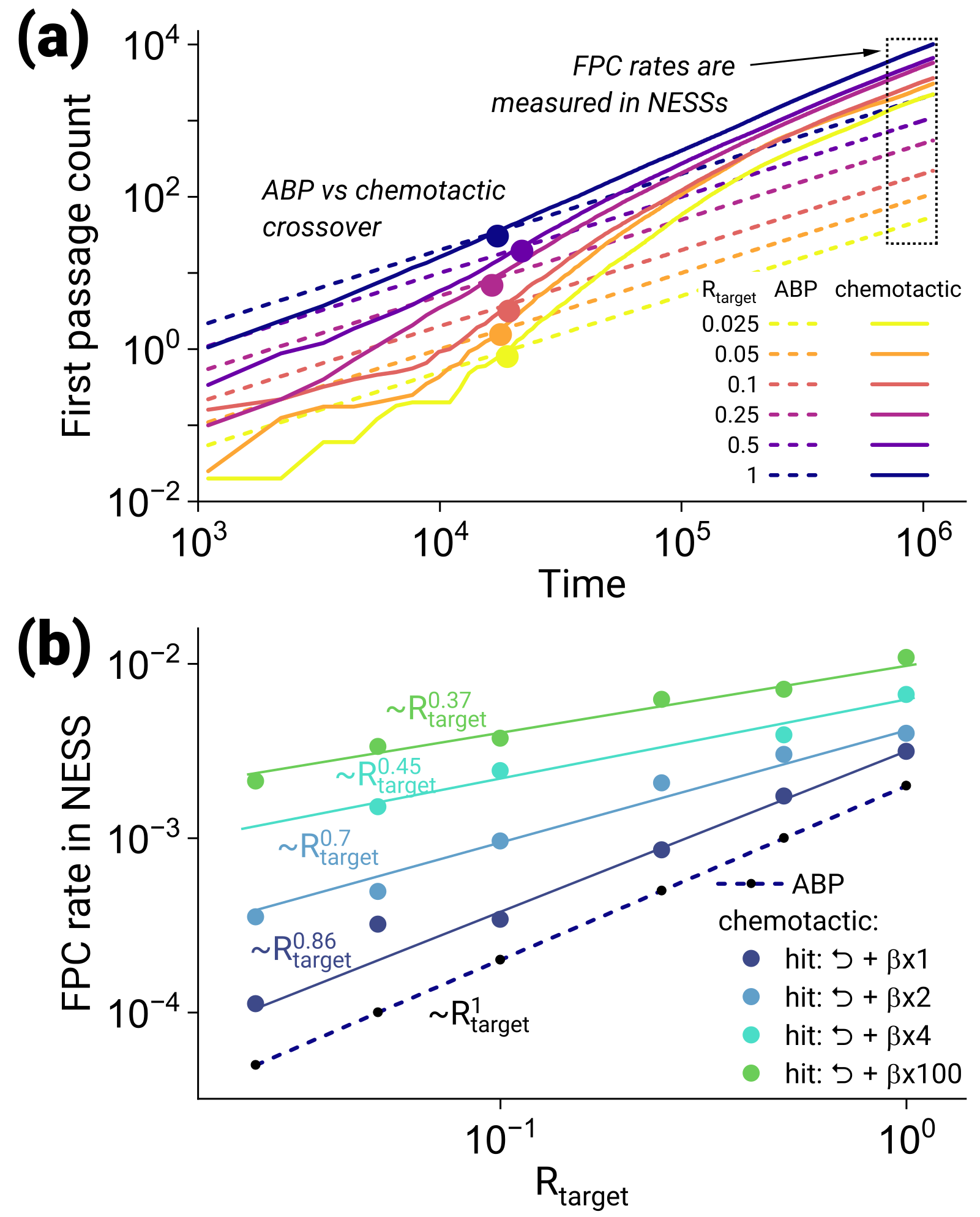}
\caption{\small{Time dependence of first passage count for pure ABPs (dashed) and chemotactic particles (solid) for different target sizes (a). First passage count rate in non-equilibrium steady state as a function of target radius $R_\text{target}$ for different target reinforcement strengths (b).}}
\label{fig:targetsize}
\end{figure}

Finally, we examined how target size influences consumption efficiency. For this, we checked targets of radii $ R_\text{target}$ from $\nicefrac{1}{80}l_p$ to $\nicefrac{1}{2}l_p$ placed  within a large circular arena ($R = 8\l_p$, $\tau_c=1.6\times10^4$) at half-radius from the arena center, $d = 4l_p$. Each system was simulated for $t_\text{sim}=1.1\times10^6$ time units, with data averaged across 50 independent realizations. First, we compared the MFPC for pure ABPs and chemotactic particles (dashed and solid lines in Fig.~\ref{fig:targetsize}a, correspondingly). As expected, smaller targets are harder to hit due to reduced cross-sectional area. Nevertheless, chemotactic search maintains a higher consumption efficiency for heavy targets compared to pure ABPs, even as the target size decreases.

To quantify sensitivity to target size, we analyzed the MFPC rate in the steady state (black frame in Fig.~\ref{fig:targetsize}a), scaling as $\sim R_\text{target}^\alpha$. Surprisingly, this sensitivity is strongly modulated by memory reinforcement (Fig.~\ref{fig:targetsize}b). For pure ABPs, the hit rate scales linearly with size ($\alpha = 1$, solid line). However, introducing even minimal reinforcement, such as a simple $180^\circ$ turn upon target contact, reduces the sensitivity to $\alpha = 0.86$. Increasing the chemoattractant deposition following a hit ($\beta_\text{return} \in [1, 100]$) further suppresses the size dependence, with scaling $\alpha$ decreasing from 0.86 to as low as 0.37.
Thus, the combination of channeling and target reinforcement not only improves consumption of heavy targets, but also significantly enhances efficiency for smaller ones.

\begin{figure}[h!]
\centering
\includegraphics[width=0.65\linewidth]{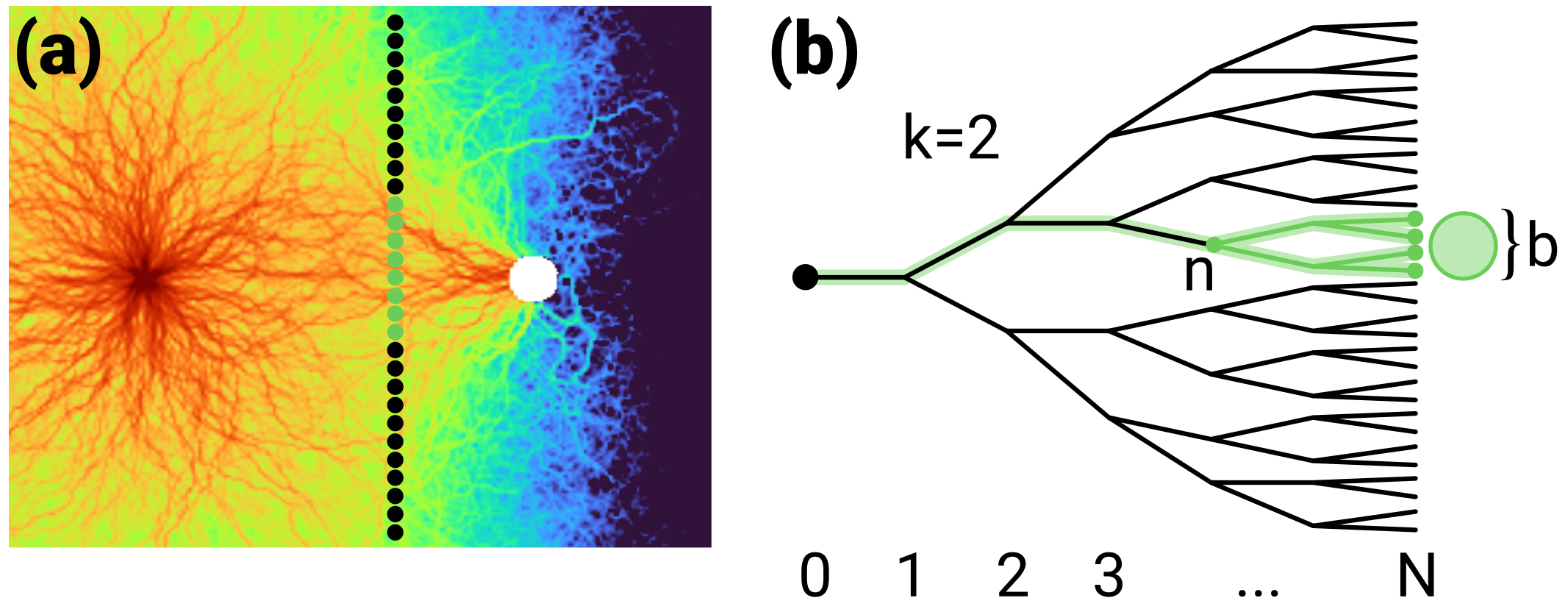}
\caption{Central part of cumulative particle density distribution $\rho$ after paths to a target are formed (a). Dots illustrate the concept of paths leading to a target (green) and missing it (black). Graphical representation of the Bethe lattice model with environmental memory (b). The green lines highlight preferred paths to the target $b$.}
\label{fig:latticemodel}
\end{figure}

The reduced sensitivity to target size resulting from reinforced target memory can be understood through a simplified analytical Bethe lattice model (see Materials and Methods). Reexamining the structure of the chemoattractant field (Fig.~\ref{fig:latticemodel}a) reveals that the frequency of target hits in steady state is determined not by the geometric cross-section of the target, but by the number of channels converging toward it. Moreover, channels directed toward the target accumulate higher chemoattractant concentrations. Based on this, we model the system as a walk on a network of channels represented by a Bethe lattice, consisting of  $N$ generations of branching points, with $k$ branches at each generation (Fig.~\ref{fig:latticemodel}b). The walker must reach a subset of $b$ branches out of a total $B = k^N $ that focus on the target (purple circles in both panels of Fig.~\ref{fig:latticemodel}).

In the absence of chemotaxis, branches are chosen randomly with equal probability $p = 1/k$ at each branching point. In this case, the probability of hitting the target is $P_0 = b/B \sim b \sim R_\text{target}$, and thus scales linearly with the target size.

However, with chemotaxis and reinforced target memory, branches leading toward the target are selected with increased probability $p_t = 1/k + \delta$, where $\delta$ reflects the chemoattractant-driven amplification. The resulting target hit probability becomes $P_t = p_t^N b^{-\log_k p_t} \sim b^\alpha$, where $\alpha = -\log_k p_t < 1$ for all $k$ and $\delta$ (see Materials and Methods). Thus, this minimal model captures the observed transition in scaling of the target hit rate with target size, from linear to sublinear, due to the effects of channeling and target memory reinforcement.

\section{Conclusions}

In this work, we analyzed the impact of chemotaxis-promoted channeling on consumption of gradually destructible targets by active Brownian particles. We demonstrated tremendous increase in mean first passage time (MFPT) in the chemotactic systems compared to pure ABPs. We demonstrated that the mean first passage count (MFPC) and it's derivative, mean first passage count rate are the relevant measures for targets requiring multiple encounters. For gradually destructible targets, these metrics determine the time to consume a target, rather than MFPT. Our model outlines the minimal requirements for optimizing MFPC via channeling: persistent motion of the searcher, alignment along preformed trails, and a target memory reinforcement mechanism. These ingredients are sufficient to break spatial symmetry in particle trajectories and drive emergence of dynamic, directional paths connecting origin to target.

Crucially, we discovered that chemotactic channeling significantly reduces sensitivity of target acquisition to target size. While target hit rates by pure ABPs exhibit linear scaling with target radius, chemotactic ABPs with reinforced memory show much weaker dependence (as low as $\sim R_\text{target}^{0.4}$), as searchers are guided straight to the target by narrow trails. This mechanism reduces variability in search outcomes, making chemotactic systems particularly effective for small and hard-to-hit targets.

Taken together, our results distinguish between two fundamentally different search paradigms: one optimized for quickly locating targets, and another for learning and reusing paths to them. While single-hit targets benefit most from the former, systems with persistent targets gain greater advantage from the latter. This distinction suggests a shift in how we analyze and design search strategies without global information, moving from the goal of faster discovery to enabling more efficient re-acquisition through environmental path learning. 
In particular, our findings provide a mechanistic rationale for the emergence of paths to persistent or gradually destructible targets in biological collectives, all from ants to large animal ecosystems, where reinforcement and environmental memory naturally enhance repeated target acquisition. Finally, these principles support broader applicability of chemotaxis-like strategies in engineered systems, offering new avenues for robust search technologies in soft active matter and swarm robotics.

\section*{Acknowledgements}
We thank Shlomi Reuveni for insightful discussions. V.R. acknowledges support from The Council of Higher Education and the Ministry of Immigration and Absorption of Israel. YR acknowledges support from the Israel Science Foundation (Grant No. 385/21) and from the European Research Council (ERC) under the European Union’s Horizon 2020 research and innovation program (Grant Agreement No. 101002392).

\section{Methods}

\subsection{Active Brownian particle with chemotaxis}

We extended the two-dimentional overdamped active Brownian particle (ABP) model by adding an interaction with environmental memory from point particle formulation of Keller-Segel model. A single particle is characterized by its position $\mathbf{r}(t)$, an intrinsic direction $\mathbf{n}(\mathbf{r})$, and an active velocity $\mathbf{v}(t)$ of a constant absolute value $|\mathbf{v}(t)|\equiv v_0$. For simplicity, we assume only rotational diffusion $D_R$, which regulates the persistence of the particle intrinsic direction $\mathbf{n}(t)$, while the translational diffusion is set to zero, $D_T=0$.

For the environmental memory, we attribute the space with a chemoattractant field $c(\mathbf{r},t)$. In the beginning of the simulations there is no predeposited chemoattractant ($c(\mathbf{r},t=0)=0$). During movement, particle deposits chemoattractant trace along its trajectory, with a Gaussian profile of intensity $\beta$ and deposition radius $r_c$. Similarly to the classical Keller-Segel model, the chemoattractant field decays exponentially over time with a half-life time $\tau_{c}$. For simplicity and computational efficiency, we do not consider the diffusion of chemoattractant.

In a feedback loop, the deposited chemoattractant affect the particles motion. We consider both translational and rotational chemotactic components to ensure simultaneously that (i) particle is dragged to a higher chemoattractant concentration areas, and (ii) particle align it's intrinsic direction with a direction perpenducular to the chemoattractant gradient (in most cases it corresponds to the established path direction, see regimes). The translational and rotational chemosensitivity constants, $\chi_T$ and $\chi_R$, respectively, regulate the intensity of these effects in the following form:

\begin{align*}
&d\mathbf{r}(t) = v_0\mathbf{\hat u}(\mathbf{r},t)\,dt\\
&\mathbf{u}(\mathbf{r},t)=\chi_T\frac{\nabla c(\mathbf{r},t)}{(c(\mathbf{r},t)+c_0)}+\mathbf{n}(t)\\
&d\varphi(t) = \left(\chi_{R} \frac{|\nabla c(\mathbf{r},t)|\cdot \sin(2\psi(t))} {\left(c(\mathbf{r,t})+c_0\right)}+ \sqrt{2 D_{R}}\eta(t)\right)dt \\
& dc(\mathbf{x},t) = \left(-\frac{1}{\tau_c}c(\mathbf{x},t)+ \beta N_{0,r_c}(|\mathbf{x}-\mathbf{r}(t)|)\right)dt
\end{align*}

where $\varphi$ represents the particle intrinsic direction,
$\mathbf{n}(t)=\{\cos\varphi(t);\sin\varphi(t)\}$; $\mathbf{\hat u}(\mathbf{r},t)=\mathbf{u}(\mathbf{r},t)/|\mathbf{u}(\mathbf{r},t)|$ characterizes the chemotaxis-induced direction of motion of a particle; $\psi(t)=\angle(\mathbf{n}(t),\nabla c(\mathbf{r},t))$ characterizes the angle between the particle intrinsic direction and chemoattractant gradient;  $c_0$ is the sensitivity noise level; $\eta(t)$ is a white noise term; and $N_{0,r_c}(r)$ is a Gaussian profile with zero mean and $\sigma=r_c$.

In this work, we used $v_0=1$, $c_0=0.001$, $r_c=0.01$. Chemosensitivities $\chi_R$, $\chi_T$ varied; for the strong chemotaxis regime, we used $\chi_R=0.15$ and $\chi_T=0.009$. Rotational diffusion $D_R$ was adjusted to obtain persistence length $l_p=1$. For $\beta$ and $\tau_c$, see below.
We solved these equations numerically using the Euler-Maruyama method with time step $\Delta t=0.005$ to ensure numerical stability. The chemoattractant distribution in the area of arena was discretized with a pixel size $\Delta x=0.001$, giving the ratio $r_c/\Delta x=10$. For computational efficiency, the chemoattractant deposition was attributed only to pixels within the cut-off radius $r_\text{cutoff}=5r_c$.

Without target reinforcement, we used $\beta=1$; for target reinforcement, see below.

\subsection{Sharp resetting}
We used sharp resetting protocol during our simulations. At $t=0$, particle started in the origin. The intrinsic direction $\mathbf{n}(t=0)$ was randomized. At a period $\tau_R$, we instantly reset the particle to the origin and randomized it's intrinsic direction again. The chemotactic field $c(\mathbf{x},t)$ was left intact under these resets until the end of the simulation. For every arena radius, we chose the reset period $\tau_R=R/v_0$, so that the particle can not escape the arena (thus we avoided boundary problems altogether).

\subsection{Target reinforcement}
For the simulations with explicit target, the particle reacted to the collision with the target (i.e. intersecting the target outer circle). The particle reaction consisted of the two changes to the particle behavior. First, in all cases, the intrinsic direction of the particle turned backwards ($\mathbf{n}\rightarrow-\mathbf{n}$). Second, in the case of target reinforcement, the chemotactic secretion $\beta$ increased temporary by $\beta_\text{mult}$ times (up to 100, see main text) and then relaxed back to $\beta=1$ exponentially with half-decay time $\tau_\beta=\tau_R$. It should be noted that even turning the intrinsic direction backwards increased the target hit rate (see Fig.\,\ref{fig:targetsize}b) and thus acted as a reinforcement. Still $\beta$ reinforcement influenced the system much stronger.

%\subsection{Mean first passage time}

\subsection{Bethe lattice model of channeling}
Let us consider Bethe lattice with branching number $k$ and $N$ generations of branches (Fig.\,\ref{fig:latticemodel}, a), which results in $k^N$ ending branches. We assume a particle travelling from the origin (left) to the end (right) by only moving to the right (no return), and equally distributed probability $1/k$ of choosing each path at all branching points. Considering a small number of branches $b$ lead to a target, one can calculate the probability to hit the target after the travel is finished: $P_0=b/k^N$. Thus the ``no environmental memory'' case leads to the linear scaling of probability of hitting the target: $P_0\sim b$. 

Now let us consider the environmental memory in the system leading to channeling, and in turn higher probabilities to choose a path to the target compared to path missing the target (Fig.\,\ref{fig:latticemodel}, b). For simplicity, let us assume that all the branches leading to the target are branching starting from generation $n$. Thus, before the branching point $n$ the probability of choosing the branch leading to the target equals $p_t=1/k+\delta$, where $\delta\geq0$ is the amplification factor due to channeling. 
Let us first express $n$ via $b$, $k$, and $N$:
$$k^{N-n}=b \Leftrightarrow n=N-\log_k b.$$

Then the probability to hit the target is the probability to follow the ``target'' path up to $n$-th branching point:
$$P_t=p_t^{n-1}=p_t^{N-1}b^{-\log_k p_t}\sim b^\alpha,$$
where $\alpha=-\log_k (1/k+\delta)$. Considering $\delta>0$, we observe a scaling factor $\alpha<1$ (for example, $\alpha\approx0.4$ for $k\in[2;20]$ and $\delta\sim0.3$.

\bibliographystyle{unsrtnat}
\bibliography{citations}  %%% Uncomment this line and comment out the ``thebibliography'' section below to use the external .bib file (using bibtex) .





% \end{thebibliography}

\end{document}